\documentclass[11pt]{article}

\usepackage[margin=1in]{geometry}
\usepackage{mathpazo}
\usepackage[T1]{fontenc}
\usepackage{amsmath,amssymb,amsthm}
\usepackage{booktabs}
\usepackage{xcolor}
\usepackage{titlesec}
\usepackage{tcolorbox}
\usepackage{authblk}
\usepackage{hyperref}
\usepackage{enumitem}
\usepackage{graphicx}
\usepackage{caption}
\usepackage[ruled,vlined,linesnumbered]{algorithm2e}

\tcbuselibrary{skins}

\SetKwInput{KwIn}{Input}
\SetKwInput{KwOut}{Output}

\definecolor{ritorange}{RGB}{247,105,2}
\definecolor{ritgray}{RGB}{88,89,91}

\titleformat{\section}
  {\normalfont\Large\bfseries\color{ritorange}}
  {\thesection}{1em}{}
\titleformat{\subsection}
  {\normalfont\large\bfseries\color{ritgray}}
  {\thesubsection}{1em}{}

\hypersetup{
  colorlinks=true,
  linkcolor=ritorange,
  citecolor=ritorange,
  urlcolor=ritorange
}

\newtcolorbox{wisdombox}{
  colback=white,
  colframe=ritorange,
  boxrule=0.8pt,
  arc=2pt,
  left=8pt, right=8pt, top=6pt, bottom=6pt,
  fontupper=\itshape\small
}

\theoremstyle{plain}
\newtheorem{proposition}{Proposition}
\theoremstyle{remark}
\newtheorem{remark}{Remark}

\title{\color{ritorange}\textbf{The Goal Attractor:\\[2pt] A Model and Test for Match-Level End Asymmetry in Association Football}}
\author{Ernest Fokoué\thanks{School of Mathematics and Statistics, Rochester Institute of Technology.}}
\date{}

\begin{document}
\maketitle

\begin{abstract}
\noindent In most football matches one goal faces more shots than the other. We ask whether that asymmetry exceeds what chance produces, and call the busier goal the \emph{Goal Attractor} and the quieter one the \emph{Goal Repeller}. Public event data are recorded relative to each team's attacking direction, so the two physical ends are not observed directly; we recover them exactly from the rule that teams change ends at half-time. We model each match's split of goalmouth events between the two ends as a beta-binomial with a latent end tilt, whose dispersion $\rho$ measures the strength of the attractor, and prove four properties: the end split is invariant to team strength and to a constant home advantage, its null distribution is exactly binomial, the overdispersion ratio is calibrated to $\rho$ and the harmonic-mean event count, and a persistent end-linked component is identified by a window covariance contrast. In 1{,}550 matches from five European leagues (StatsBomb Open Data, 2015/16), the busier goal receives 59.8\% of shots against 58.2\% expected by chance; the overdispersion ratio is 1.40 (bootstrap 95\% CI $[1.31,\,1.49]$, $\hat\rho=0.017$), it replicates in four leagues and holds with the score level. The dispersion is similar for shots on target and clear chances ($\hat\rho=0.012$--$0.021$) and not detectable for goals, which are too few to test. A persistent end effect large enough to explain the excess is rejected; modest match-specific variation in relative attacking intensity reproduces it. The asymmetry is small and not usefully predictable within a match.

\end{abstract}

\section{Introduction}

Watch enough football and one regularity is hard to miss: in a given match, one goal seems to live under siege while the other is left in relative peace, and the busy end is not always the same end. Counting shots at each end confirms that nearly every match has a busier goal, but that observation carries little information by itself, because chance alone splits a finite number of shots unevenly. The substantive question is whether the split is more uneven, across matches, than chance allows. We call the goal that faces more goalmouth action in a match the \emph{Goal Attractor} and the other the \emph{Goal Repeller}, and ask:
\begin{quote}
\textbf{$H_0$:} in every match, shots are allocated to the two ends as fair coin flips, conditional on their number.\\[2pt]
\textbf{$H_a$:} some matches have a genuine end tilt, so the split varies across matches by more than chance.
\end{quote}
Three obstacles make this harder than it sounds. First, event data are recorded relative to each team's own attacking direction, so the physical ends of the ground are never observed. Second, the attractor is defined within a match, so a test must be invariant to which team is stronger and to how large home advantage is. Third, a rejection of $H_0$ is not by itself evidence of a physical, end-linked cause, and one needs to know what else could produce it. This paper addresses each in turn. Section~2 shows how to recover the two ends exactly from the half-time swap, specifies a latent-tilt model, and proves four properties of the resulting statistic. Section~3 applies the test to 1{,}550 matches and examines the score state, the type of event counted, the independence of the result from a persistent end effect, and its practical predictability. Section~4 discusses what is and is not established.

\subsection{Related work}
Home advantage is among the most robust findings in sports science (Pollard 1986), and recent work asks whether it is constant within a match. Price, Cai, Shen, and Hu (2022) used causal inference on English Premier League data to quantify its size, and Qi, Cai, Hu, and Shen (2025) estimate it as a time-varying, non-stationary causal process using expected-goals and event data. These studies are relevant here for a precise reason: Proposition~\ref{prop:structure} below shows that the end split responds to exactly one kind of variation, a change between the two halves in how far the home team's attacking rate exceeds the visitors', and is silent about a constant home advantage. Spatial shot-chart modelling in basketball (Reich, Hodges, Carlin, and Reich 2006; Yamada and Fujii 2026) treats location asymmetry through hierarchical random effects, which is the modelling idea we adapt, with the difference that football's ends swap at half-time and so must be recovered rather than read from fixed court geometry.

\section{Setting, Model and Theory}

\subsection{Data}
We use shot-event data from StatsBomb Open Data (StatsBomb 2024) for five European top-flight leagues, all 2015/16: La Liga (379 usable matches, 9{,}152 shots), the Premier League (380 matches, 9{,}908 shots), Serie A (380, 9{,}998), Ligue 1 (377, 8{,}814) and Bundesliga, for which the open release covers only 34 matches (840 shots), giving 1{,}550 matches and 38{,}712 shots in all. A match is usable when both teams have recorded shots. The first four are complete seasons; Bundesliga is reported for completeness and is underpowered. An independent dataset, the FA Women's Super League 2020/21, is analysed in Section~3.6. Shots in extra time are excluded. For outcome-specific analyses (Section~3.3) we also use each shot's recorded outcome and StatsBomb's expected-goals value (\texttt{statsbomb\_xg}).

\subsection{Recovering the two ends}
\begin{figure}[h]
\centering
\includegraphics[width=0.7\textwidth]{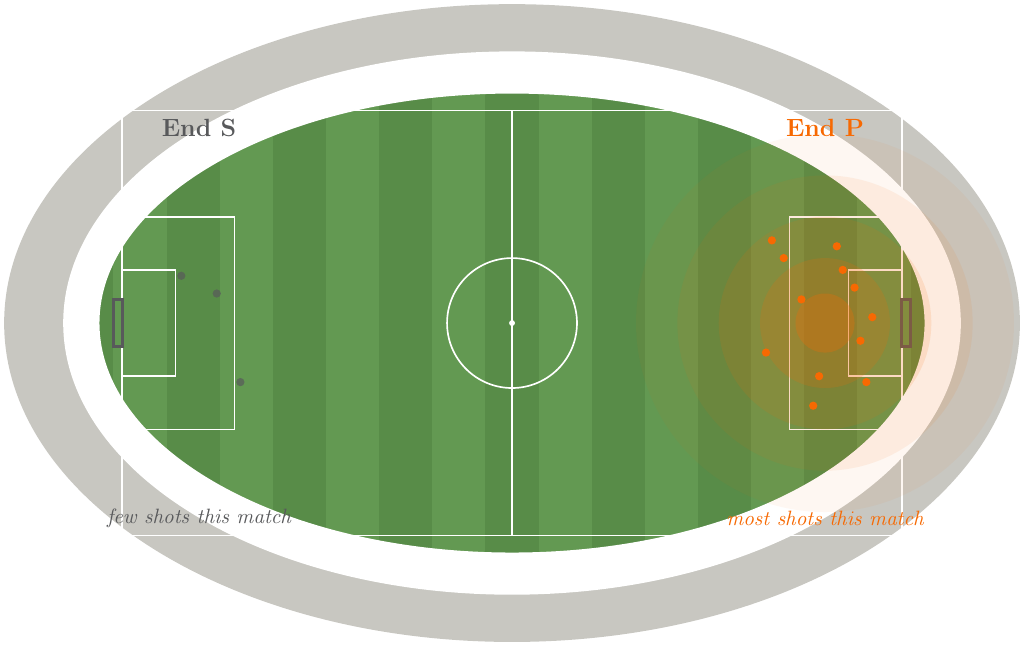}
\caption{Schematic of the Goal Attractor. In this illustrative match one end (here End~P) draws most of the shots; the other (End~S) is the Goal Repeller. Which physical end plays each role can differ from match to match.}
\label{fig:stadium}
\end{figure}

Let $s(T,k)$ be the number of events (shots, unless stated) taken by team $T$ in period $k\in\{1,2\}$, for home team $H$ and away team $A$. Teams change ends at half-time, so the team that attacks one physical goal in the first half attacks the other in the second. Hence the events directed at one physical goal are those taken by $H$ in period 1 and by $A$ in period 2, and the events directed at the other are the remainder:
\begin{equation}
\text{End}_P = s(H,1) + s(A,2), \qquad \text{End}_S = s(A,1) + s(H,2).
\label{eq:endswap}
\end{equation}
This is an exact partition of the match's events by physical end, with the labelling fixed by convention ($P$ is the end the home team attacks first). The labelling carries no information about the ground, which is why the method needs no stadium geometry. Figure~\ref{fig:stadium} illustrates the idea and Algorithm~\ref{alg:endswap} gives the computation. The \emph{Goal Attractor} of a match is the end with the larger count; it can be either $P$ or $S$.

\begin{algorithm}[H]
\caption{End assignment for a single match}
\label{alg:endswap}
\KwIn{Event log for one match; home team $H$; away team $A$}
\KwOut{$\text{End}_P$, $\text{End}_S$}
$\text{End}_P \gets 0$;\ $\text{End}_S \gets 0$\;
\ForEach{shot $s$ in the match (periods 1 and 2)}{
  \uIf{$\mathrm{period}(s) = 1$}{
    \uIf{$\mathrm{team}(s) = H$}{$\text{End}_P \gets \text{End}_P + 1$}
    \Else{$\text{End}_S \gets \text{End}_S + 1$}
  }
  \Else{
    \tcp{second half: the teams have swapped ends}
    \uIf{$\mathrm{team}(s) = H$}{$\text{End}_S \gets \text{End}_S + 1$}
    \Else{$\text{End}_P \gets \text{End}_P + 1$}
  }
}
\Return{$\text{End}_P$, $\text{End}_S$}
\end{algorithm}

\subsection{The Goal Attractor model}
Let match $i$ have $n_i=\text{End}_{P,i}+\text{End}_{S,i}$ events, and let $X_i=\text{End}_{P,i}$. We posit a latent end tilt $\pi_i\in(0,1)$, the probability that an event in match $i$ falls at End~P, with
\begin{equation}
X_i\mid \pi_i \sim \mathrm{Binomial}(n_i,\pi_i),\qquad \pi_i\sim\mathrm{Beta}(a,a),\qquad a=\frac{1-\rho}{2\rho}.
\label{eq:model}
\end{equation}
The tilt is symmetric about $1/2$ because the labels $P$ and $S$ are arbitrary. Marginally $X_i$ is beta-binomial with mean $n_i/2$ and intra-match dispersion $\rho\in[0,1]$, where $\rho=0$ gives $\pi_i\equiv 1/2$ and $H_0$. The \emph{strength of the Goal Attractor} is $\rho$, and since $\operatorname{Var}(\pi_i)=\rho/4$ it has a direct reading: the attractor's true share of events in match $i$ is $\max(\pi_i,1-\pi_i)$, whose average is about $\tfrac12+0.4\sqrt{\rho}$ for small $\rho$. For $\rho=0.0174$ (the pooled estimate for shots, Section~3.1), the average true attractor share is 55.3\%, and the standard deviation of $\pi_i$ is $0.066$.

\begin{remark}[Chance alone creates a busier goal]\label{rem:chance}
Even with $\rho=0$ one end almost always receives more events, so the observed attractor share $\max(\hat p_i,1-\hat p_i)$ is above $1/2$ for every match with a non-tied split. With roughly 25 shots per match its chance expectation is about 58\%, not 50\%. The relevant comparison is therefore between the observed and chance-expected attractor share, or equivalently between the observed and chance-expected variance of $\hat p_i=X_i/n_i$, which is the approach taken below.
\end{remark}

\subsection{Theory}
Let $D_k=s(H,k)-s(A,k)$ be the home-minus-away event differential in period $k$.

\begin{proposition}[Structure of the end split]\label{prop:structure}
(a) For every match, $\text{End}_P-\text{End}_S = D_1-D_2$.\\
(b) Suppose that, given the match, the four counts $s(H,1),s(A,1),s(H,2),s(A,2)$ are independent Poisson variables with means $\lambda_{H1},\lambda_{A1},\lambda_{H2},\lambda_{A2}$. Put $\Delta_k=\lambda_{Hk}-\lambda_{Ak}$ and $\Lambda=\lambda_{H1}+\lambda_{A1}+\lambda_{H2}+\lambda_{A2}$. Then, conditional on $n$, $\text{End}_P\sim\mathrm{Binomial}(n,\pi)$ with
\begin{equation}
\pi=\frac12+\frac{\Delta_1-\Delta_2}{2\Lambda}.
\label{eq:pi}
\end{equation}
In particular $\pi=1/2$ exactly when the home-minus-away rate difference is the same in the two halves, whatever the teams' strengths, whatever the size of a constant home advantage, and whatever the overall tempo.
\end{proposition}
\begin{proof}
(a) Expand: $[s(H,1)+s(A,2)]-[s(A,1)+s(H,2)]=[s(H,1)-s(A,1)]-[s(H,2)-s(A,2)]$. (b) By independence and the superposition property of Poisson variables (Kingman 1993), $\text{End}_P\sim\mathrm{Poisson}(\lambda_{H1}+\lambda_{A2})$ and $\text{End}_S\sim\mathrm{Poisson}(\lambda_{A1}+\lambda_{H2})$ are independent. Conditional on their sum $n$, $\text{End}_P$ is binomial with success probability $(\lambda_{H1}+\lambda_{A2})/\Lambda$. Writing $\lambda_{H1}+\lambda_{A2}=\tfrac12\Lambda+\tfrac12(\Delta_1-\Delta_2)$ gives \eqref{eq:pi}.
\end{proof}
Part (a) shows that the end contrast is the first-half minus second-half home-versus-away differential; part (b) shows why this is not a home-advantage statistic: a home advantage that is the same in both halves enters $\Delta_1$ and $\Delta_2$ equally and cancels. What the end split detects is any systematic difference between the halves in the relative attacking rates of the two sides. A physical end effect (a goal that is busier for both teams, in both halves) is one such source, since it adds to $\Delta_1$ and subtracts from $\Delta_2$. A half-specific change in relative intensity (tactics, substitutions, fatigue, game state) is another, and Proposition~\ref{prop:identify} shows how the two can be partly separated.

\begin{proposition}[Exact null distribution]\label{prop:null}
If $\Delta_1=\Delta_2$ in every match, then, conditional on $(n_i)_{i=1}^M$, the counts $X_i$ are independent $\mathrm{Binomial}(n_i,1/2)$. Consequently the permutation test of Section~2.5, which resamples $X_i$ from this distribution, has exact level.
\end{proposition}
\begin{proof}
Immediate from Proposition~\ref{prop:structure}(b) with $\pi=1/2$ and independence across matches.
\end{proof}

\begin{proposition}[Calibration of the overdispersion ratio]\label{prop:calib}
Under the model~\eqref{eq:model}, $\operatorname{Var}(X_i/n_i)=\frac{1}{4n_i}\big[1+(n_i-1)\rho\big]$. With $V_{\mathrm{obs}}$ the sample variance of $\hat p_i$ and $R=V_{\mathrm{obs}}/\bar V_{\mathrm{exp}}$ (Section~2.5),
\begin{equation}
\mathbb{E}[R]\;\approx\;1+\rho\,(n_h-1),\qquad n_h=\Big(\tfrac1M\textstyle\sum_i 1/n_i\Big)^{-1},
\label{eq:calib}
\end{equation}
up to terms of order $1/M$, where $n_h$ is the harmonic-mean event count.
\end{proposition}
\begin{proof}
The variance of a beta-binomial proportion with mean $1/2$ is $\frac{1}{4n}[1+(n-1)\rho]$ (Johnson, Kemp, and Kotz 2005). Averaging over matches and dividing by $\bar V_{\mathrm{exp}}=\frac1M\sum_i\frac{1}{4n_i}$ gives $1+\rho\,\frac{\frac1M\sum_i (n_i-1)/n_i}{\frac1M\sum_i 1/n_i}=1+\rho(n_h-1)$.
\end{proof}
Equation~\eqref{eq:calib} matters for interpretation: $R$ depends on how many events are counted per match, so the same end tilt produces a large $R$ for shots ($n_h\approx 24$) and a small one for goals ($n_h\approx 2$). Comparisons across event types should therefore be made in $\hat\rho$, not in $R$.

\begin{proposition}[A persistent end-linked component is identified]\label{prop:identify}
Split each half into an early and a late window, giving four windows $w\in\{1a,1b,2a,2b\}$ and window differentials $D_w=m+s_w\delta+\eta_w+\varepsilon_w$ with orientation signs $s=(+1,+1,-1,-1)$. Here $m$ is the match's persistent home-versus-away level (team strengths and home advantage), $\delta$ is a persistent end effect, $\eta_w$ is a short-memory window shock, and $\varepsilon_w$ is Poisson sampling noise, with $\delta$ independent of $(m,\eta)$ and $\varepsilon$ uncorrelated across windows. Let $c_w=\operatorname{Cov}(\eta_{1a},\eta_{1b})=\operatorname{Cov}(\eta_{2a},\eta_{2b})$ and $c_x=\operatorname{Cov}(\eta_{1b},\eta_{2a})$. Then
\begin{equation}
T=\tfrac12\big[\operatorname{Cov}(D_{1a},D_{1b})+\operatorname{Cov}(D_{2a},D_{2b})\big]-\operatorname{Cov}(D_{1b},D_{2a})=2\operatorname{Var}(\delta)+(c_w-c_x),
\label{eq:T}
\end{equation}
and, with $\Delta=D_1-D_2$,
\begin{equation}
2\operatorname{Var}(\delta)\;\le\;U:=\tfrac18\big(\operatorname{Var}(\Delta)-\mathbb{E}[n]\big).
\label{eq:U}
\end{equation}
If halftime does not strengthen the persistence of window shocks ($c_w\ge c_x$), then $T$ is an upper bound on $2\operatorname{Var}(\delta)$.
\end{proposition}
\begin{proof}
The terms $\operatorname{Var}(m)$ cancel in the difference defining $T$ because $m$ enters every window with sign $+1$. The within-half covariance is $\operatorname{Var}(m)+\operatorname{Var}(\delta)+c_w$ (the two windows share the sign of $\delta$), while the covariance across half-time is $\operatorname{Var}(m)-\operatorname{Var}(\delta)+c_x$ (opposite signs), which gives \eqref{eq:T}. Poisson noise contributes nothing to covariances between distinct windows. For \eqref{eq:U}: $\Delta=\sum_w s_w D_w=4\delta+\sum_w s_w\eta_w+\sum_w s_w\varepsilon_w$ because $\sum_w s_w=0$, so $\operatorname{Var}(\Delta)=16\operatorname{Var}(\delta)+\operatorname{Var}(\sum_w s_w\eta_w)+\mathbb{E}[n]$, the last term being the Poisson variance of a difference of counts. Dropping the non-negative middle term gives $16\operatorname{Var}(\delta)\le\operatorname{Var}(\Delta)-\mathbb{E}[n]$.
\end{proof}
Here $U$ is the largest value of $2\operatorname{Var}(\delta)$ compatible with the observed spread of $\Delta$, that is, the value it would take if \emph{all} of the excess dispersion were a persistent end effect. Comparing $T$ with $U$ therefore says how much of the excess such an effect can account for.

\begin{remark}[Dispersion without any end effect]\label{rem:frailty}
Suppose there is no end effect at all, but each team's attacking intensity in each half is multiplied by an independent Gamma factor with mean 1 and variance $\kappa$ (a coefficient of variation $\sqrt\kappa$). Conditional on $n$, $\text{End}_P$ is then overdispersed. In a simulation with six expected shots per team-half, $R=1.27$ at $\kappa=0.05$ and $R=1.49$ at $\kappa=0.10$, so the observed $R=1.40$ corresponds to $\kappa\approx0.08$, a coefficient of variation of about 0.28 in a team's half-by-half intensity. Excess dispersion is thus what ordinary within-match variation in relative attacking intensity produces, and the test cannot by itself attribute it to a physical end.
\end{remark}

\subsection{Inference}
Let $\hat p_i=X_i/n_i$ and $V_{\mathrm{obs}}=\operatorname{Var}_i(\hat p_i)$. Under $H_0$ and conditional on $n_i$, $\operatorname{Var}(\hat p_i\mid n_i)=0.25/n_i$ (Proposition~\ref{prop:null}), so the chance-expected variance is
\begin{equation}
\bar V_{\text{exp}} = \frac{1}{M}\sum_{i=1}^{M} \frac{0.25}{n_i},\qquad R = \frac{V_{\text{obs}}}{\bar V_{\text{exp}}}.
\label{eq:ratio}
\end{equation}
Significance uses a permutation test in the tradition of Good (2005): we draw $B$ synthetic samples $\tilde x_i\sim\mathrm{Binomial}(n_i,0.5)$ matched on each match's observed $n_i$ and compute
\begin{equation}
p = \frac{1}{B}\sum_{b=1}^{B} \mathbb{1}\!\left[V^{(b)}_{\text{sim}} \geq V_{\text{obs}}\right].
\label{eq:permp}
\end{equation}
(Algorithm~\ref{alg:permtest}). Confidence intervals for $R$ and $\rho$ come from resampling matches with replacement (Efron and Tibshirani 1993), and $\hat\rho$ is the maximum-likelihood estimate of the beta-binomial model~\eqref{eq:model} with mean fixed at $1/2$. The test is applied to any count that can be assigned to ends by Algorithm~\ref{alg:endswap}: all shots, shots on target, clear chances, or goals.

\begin{algorithm}[H]
\caption{Overdispersion permutation test}
\label{alg:permtest}
\KwIn{Pairs $\{(X_i,\,n_i)\}_{i=1}^{M}$ from Algorithm~\ref{alg:endswap}; number of permutations $B$}
\KwOut{Overdispersion ratio $R$; permutation $p$-value}
\ForEach{$i = 1,\dots,M$}{$\hat p_i \gets X_i/n_i$}
$V_{\text{obs}} \gets \mathrm{Var}(\hat p_1,\dots,\hat p_M)$\;
$\bar V_{\text{exp}} \gets \frac{1}{M}\sum_{i=1}^{M} 0.25/n_i$\;
$R \gets V_{\text{obs}} / \bar V_{\text{exp}}$;\ $\text{count} \gets 0$\;
\For{$b = 1$ \KwTo $B$}{
  \ForEach{$i = 1,\dots,M$}{draw $x_i \sim \mathrm{Binomial}(n_i,\,0.5)$;\ $\tilde p_i \gets x_i / n_i$\;}
  $V^{(b)}_{\text{sim}} \gets \mathrm{Var}(\tilde p_1,\dots,\tilde p_M)$\;
  \If{$V^{(b)}_{\text{sim}} \geq V_{\text{obs}}$}{$\text{count} \gets \text{count} + 1$}
}
$p \gets \text{count}/B$\;
\Return{$R$, $p$}
\end{algorithm}

\paragraph{Persistent-end contrast.} For Proposition~\ref{prop:identify} we split the first half at minute 23 and the second at minute 68 (stoppage time falls in the late windows), compute $D_w$ for each match, and estimate $T$ and $U$ by their sample analogues, with bootstrap intervals from resampling matches (Algorithm~\ref{alg:persist}).

\begin{algorithm}[H]
\caption{Persistent-end contrast}
\label{alg:persist}
\KwIn{Per-match window differentials $D_{1a},D_{1b},D_{2a},D_{2b}$ and shot totals $n_i$; bootstrap replicates $S$}
\KwOut{$\hat T$, $\hat U$ with 95\% bootstrap intervals}
\For{replicate $=0,1,\dots,S$ (replicate 0 uses the data; others resample matches)}{
  $\hat T \gets \tfrac12[\widehat{\mathrm{Cov}}(D_{1a},D_{1b})+\widehat{\mathrm{Cov}}(D_{2a},D_{2b})]-\widehat{\mathrm{Cov}}(D_{1b},D_{2a})$\;
  $\Delta_i \gets D_{1a,i}+D_{1b,i}-D_{2a,i}-D_{2b,i}$;\ $\hat U \gets \tfrac18[\widehat{\mathrm{Var}}(\Delta)-\bar n]$\;
}
\Return{$\hat T,\hat U$ and percentile intervals}
\end{algorithm}

\paragraph{Power.} To judge what a non-significant result can mean, Algorithm~\ref{alg:power} simulates samples with the sample's own event counts and a hypothesised true ratio $R^\star$, then applies Algorithm~\ref{alg:permtest}. Setting $\operatorname{sd}_i^2=(R^\star-1)\cdot0.25/n_i$ makes the simulated ratio equal $R^\star$ in expectation.

\begin{algorithm}[H]
\caption{Power simulation}
\label{alg:power}
\KwIn{Event totals $\{n_i\}_{i=1}^{M}$; hypothesised ratio $R^\star$; replicates $S$; permutations $B$}
\KwOut{Estimated statistical power}
$\text{rejections} \gets 0$\;
\For{$s = 1$ \KwTo $S$}{
  \ForEach{$i = 1,\dots,M$}{
    $\text{sd}_i \gets \sqrt{\max\big((R^\star-1)\cdot 0.25/n_i,\ 0\big)}$;\ draw $p_i^\star \sim \mathcal{N}(0.5,\,\text{sd}_i^2)$, clipped to $(0.01,\,0.99)$\;
    draw $x_i \sim \mathrm{Binomial}(n_i,\,p_i^\star)$\;
  }
  run Algorithm~\ref{alg:permtest} on $\{(x_i,\,n_i)\}$ with $B$ permutations to get $p^{(s)}$\;
  \If{$p^{(s)} < 0.05$}{$\text{rejections} \gets \text{rejections} + 1$}
}
\Return{$\text{rejections}/S$}
\end{algorithm}

\section{Results}

\subsection{Primary test and cross-league replication}
\begin{table}[h]
\centering\small
\begin{tabular}{@{}lrcccc@{}}
\toprule
Sample & $n$ & $R$ & Bootstrap 95\% CI & Perm.\ $p$ & Attractor share (obs./chance) \\
\midrule
La Liga 2015/16        & 379 & 1.492 & [1.290, 1.698] & $<0.001$ & 0.604 / 0.583 \\
Premier League 2015/16 & 380 & 1.326 & [1.135, 1.515] & $<0.001$ & 0.592 / 0.580 \\
Serie A 2015/16        & 380 & 1.545 & [1.340, 1.751] & $<0.001$ & 0.600 / 0.579 \\
Ligue 1 2015/16        & 377 & 1.267 & [1.100, 1.425] & $<0.001$ & 0.595 / 0.584 \\
Bundesliga 2015/16*    &  34 & 1.154 & [0.698, 1.582] & 0.21 & 0.593 / 0.582 \\
\midrule
Pooled, five leagues   & 1550 & 1.402 & [1.312, 1.494] & $<0.001$ & 0.598 / 0.582 \\
\bottomrule
\end{tabular}
\caption{Overdispersion of the end split for shots. ``Attractor share'' is the mean over matches of $\max(\hat p_i,1-\hat p_i)$, observed and as expected under $H_0$ (mean of 1{,}000 chance samples). *Bundesliga's open release covers only 34 matches; its estimate points the same way but is underpowered.}
\label{tab:main}
\end{table}

\begin{figure}[h]
\centering
\includegraphics[width=0.78\textwidth]{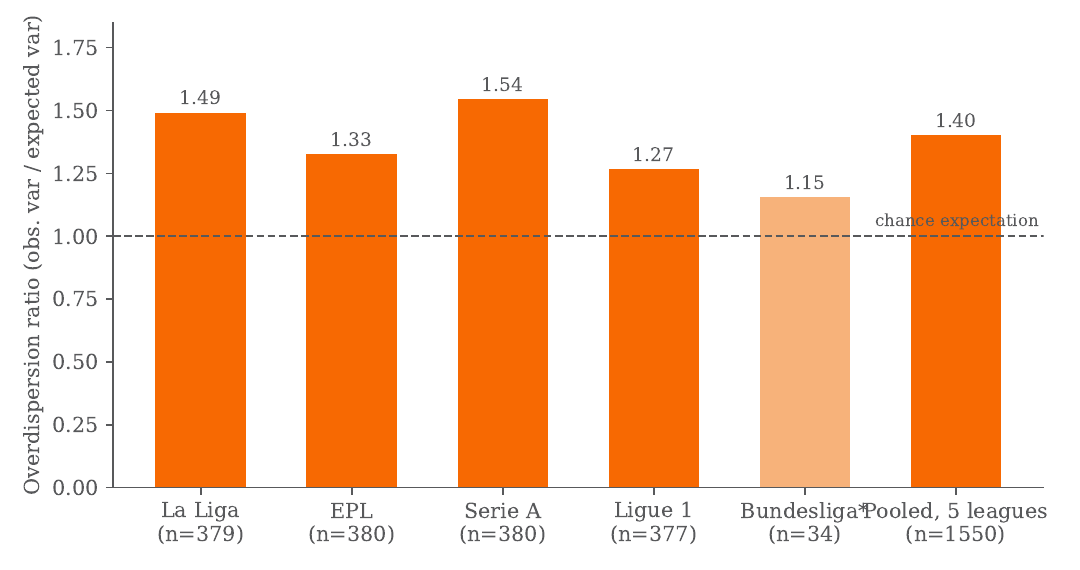}
\caption{Overdispersion ratio $R$ for shots, five leagues and pooled. The dashed line marks the chance expectation ($R=1$). The muted bar is the underpowered Bundesliga sample.}
\label{fig:ratios}
\end{figure}

The busier goal receives 59.8\% of shots against 58.2\% expected by chance (Remark~\ref{rem:chance}); the excess is small in absolute terms but far outside chance variation (Table~\ref{tab:main}, Figure~\ref{fig:ratios}). The estimate replicates independently in four leagues with no shared teams, venues or officials. By Proposition~\ref{prop:calib}, the pooled $R=1.40$ with $n_h=23.5$ corresponds to $\hat\rho=0.0174$ (bootstrap 95\% CI $[0.0135,\,0.0210]$), that is, a typical attractor share of about 55\% in the underlying tilt.

\paragraph{What a non-significant Bundesliga result means.} Using the 34 matches' own shot counts and $R^\star=1.40$, Algorithm~\ref{alg:power} (3{,}000 replicates) gives power $0.41$: even if Bundesliga's true effect equalled the pooled one, it would be detected only about two times in five. The observed $p=0.21$ is therefore consistent with underpowering alone.

\begin{figure}[h]
\centering
\includegraphics[width=0.78\textwidth]{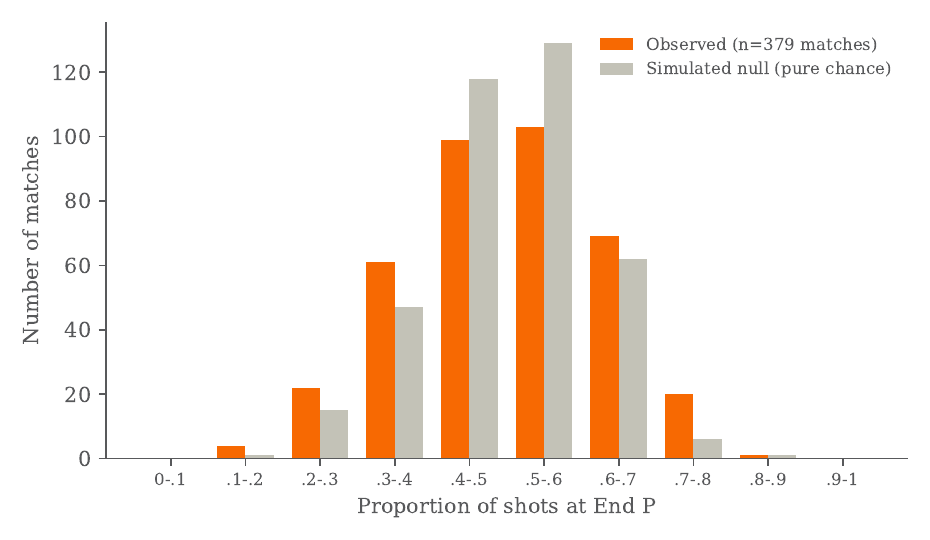}
\caption{Distribution of the per-match share of shots at End~P, La Liga 2015/16 ($n=379$). Observed matches (orange) show heavier tails than a chance simulation (gray).}
\label{fig:histogram}
\end{figure}

\subsection{Score state}
A losing team pushes forward, which concentrates play near one goal; the end split could therefore reflect score-chasing. Table~\ref{tab:score} gives two controls: matches that are level at half-time (so no first-half score effect), and counts restricted to shots taken while the score was exactly level, which removes score-chasing at every moment of the match. Goals are tracked from shot outcomes; own goals (about 1\% of goals) are not.

\begin{table}[h]
\centering\small
\begin{tabular}{@{}lccc@{}}
\toprule
Sample & All matches & Level at half-time & Shots while score level \\
\midrule
La Liga      & 1.492 (379) & 1.245 (146), $p=0.020$ & 1.608 (272), $p<0.001$ \\
Premier League & 1.326 (380) & 1.315 (177), $p=0.002$ & 1.426 (293), $p<0.001$ \\
Serie A      & 1.545 (380) & 1.224 (168), $p=0.021$ & 1.704 (297), $p<0.001$ \\
Ligue 1      & 1.267 (377) & 1.310 (167), $p=0.004$ & 1.281 (293), $p=0.001$ \\
Bundesliga*  & 1.154 (34)  & 1.167 (14), $p=0.22$ & 1.425 (24), $p=0.061$ \\
\midrule
Pooled       & 1.402 (1550) & 1.283 (672), $p<0.001$ & 1.509 (1179), $p<0.001$ \\
\bottomrule
\end{tabular}
\caption{Overdispersion ratio $R$ for shots (matches in parentheses) under two score-state controls. Matches with fewer than six qualifying shots are dropped.}
\label{tab:score}
\end{table}

The effect survives both controls. It is smaller in matches level at half-time ($R=1.28$) and larger when only level-score shots are counted ($R=1.51$), so live score-chasing does not generate it. We also regressed the raw contrast $\Delta_i=\text{End}_{P,i}-\text{End}_{S,i}$ on the match's home-minus-away total shot differential and its second-half-minus-first-half total shot differential, which are two natural nuisance quantities; together they explain only $R^2=0.070$ of the variance of $\Delta_i$.

\subsection{Shots, shots on target, clear chances and goals}
The question that motivates the paper concerns where \emph{dangerous} play concentrates, so we repeat the analysis for three stricter counts: shots on target (goals, saves, saves onto the post), clear chances (shots with $xG\ge0.1$), and goals.

\begin{table}[h]
\centering\scriptsize
\resizebox{\textwidth}{!}{\begin{tabular}{@{}lrrrcccc@{}}
\toprule
Count & Matches & Events & $n_h$ & $R$ [95\% CI] & $R$ predicted at $\hat\rho=0.0174$ & $p$ & $\hat\rho$ [95\% CI] \\
\midrule
Shots          & 1550 & 38{,}712 & 23.5 & 1.402 [1.312, 1.494] & 1.40  & $<0.001$ & 0.0174 [0.0135, 0.0210] \\
On target      & 1550 & 13{,}082 &  7.1 & 1.104 [1.023, 1.181] & 1.107 & 0.002    & 0.0117 [0.0015, 0.0214] \\
Clear chances  & 1545 &  9{,}861 &  4.8 & 1.116 [1.050, 1.178] & 1.066 & $<0.001$ & 0.0209 [0.0081, 0.0323] \\
Goals          & 1413 &  3{,}958 &  2.1 & 0.993 [0.955, 1.030] & 1.018 & 0.63     & 0.0008 [0.0000, 0.0269] \\
\bottomrule
\end{tabular}}
\caption{Pooled five-league results by type of event counted. The column ``$R$ predicted'' applies Proposition~\ref{prop:calib} with the shots estimate $\hat\rho=0.0174$ and each count's own harmonic-mean size $n_h$.}
\label{tab:outcomes}
\end{table}

\begin{figure}[h]
\centering
\includegraphics[width=0.72\textwidth]{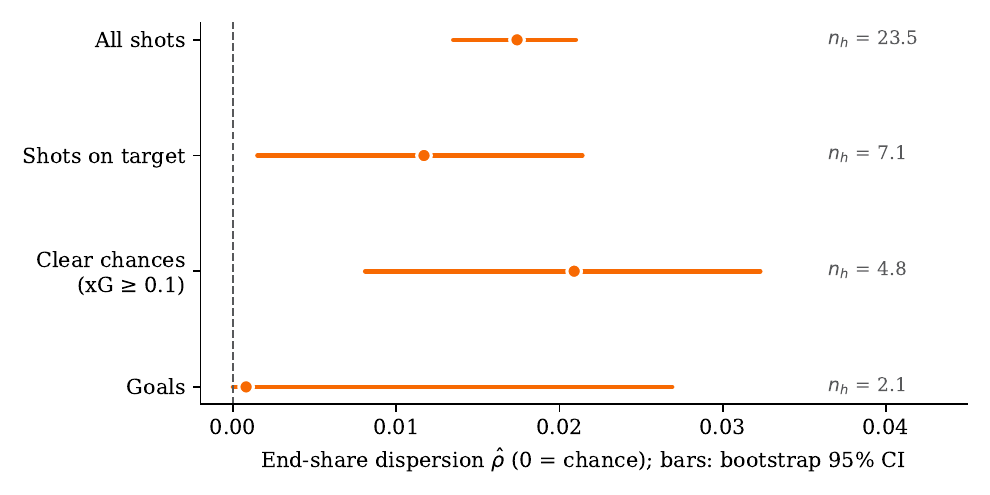}
\caption{End-share dispersion $\hat\rho$ by type of event counted, pooled over five leagues. Zero is chance. Intervals are bootstrap 95\%.}
\label{fig:outcomes}
\end{figure}

The raw ratio falls from 1.40 for shots to 1.10 for shots on target and 1.12 for clear chances, and is indistinguishable from 1 for goals (Table~\ref{tab:outcomes}). This pattern is largely what Proposition~\ref{prop:calib} predicts for a constant end tilt: because $R=1+\rho(n_h-1)$, an unchanged $\rho$ gives a smaller $R$ when fewer events are counted. The tilt itself is estimated at similar sizes for the three shot-based counts ($\hat\rho=0.012$ to $0.021$, with overlapping intervals; Figure~\ref{fig:outcomes}). For goals the evidence is uninformative rather than negative: with 2.1 goals per match, the ratio expected under $\rho=0.0174$ is only $1.018$, and the observed $0.993$ (CI $[0.955,\,1.030]$) is compatible both with that value and with none. Goals are too sparse for this design to detect an effect of this size, and we make no claim about them either way.

\subsection{Is the excess dispersion a persistent end effect?}
The Goal Attractor hypothesis in its strongest form says that the same physical goal stays busy for the whole match. Proposition~\ref{prop:identify} turns this into a testable contrast. Using four windows per match (minutes 0--23 and 23--45 of the first half; 45--68 and 68 onward in the second), the pooled estimates are $\hat T=0.07$ shots$^2$ (bootstrap 95\% CI $[-0.47,\,0.62]$), against $\hat U=1.37$ ($[1.07,\,1.69]$), the value $T$ would take if the whole excess dispersion were a persistent end effect. The within-half covariance of adjacent windows (1.67) is almost identical to the covariance across half-time (1.61), so there is no sign of the sign-flipped persistence a physical end effect would produce. The difference $\hat T-\hat U=-1.30$ has bootstrap 95\% interval $[-1.75,\,-0.84]$, so a persistent end effect large enough to explain the excess is rejected, and the upper confidence limit for $T$ is under half of $\hat U$. Under the one-sided assumption $c_w\ge c_x$, $\operatorname{sd}(\delta)\le\sqrt{0.62/2}\approx0.56$ shots per window, small relative to windows that carry about six shots. By Remark~\ref{rem:frailty}, a coefficient of variation of about 0.28 in a team's half-by-half intensity would produce the observed dispersion with no end effect at all.

\begin{figure}[h]
\centering
\includegraphics[width=0.78\textwidth]{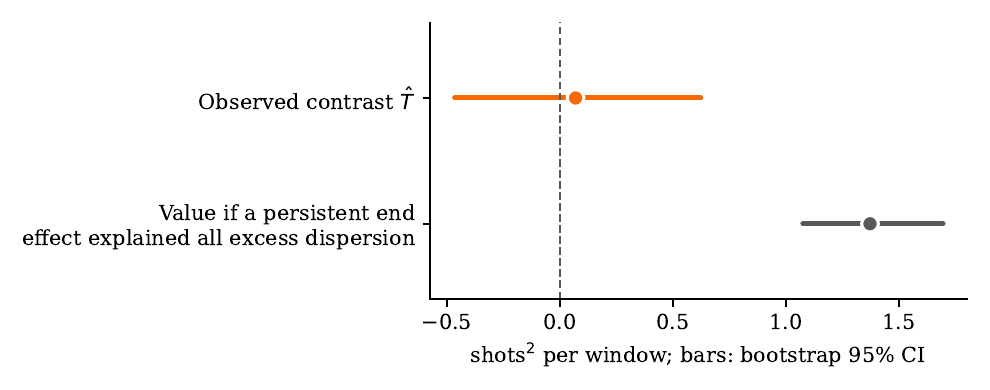}
\caption{Persistent-end contrast. Orange: the observed $\hat T$. Gray: the value $\hat U$ it would take if a persistent end effect explained all excess dispersion. Pooled over four leagues plus Bundesliga ($n=1550$); intervals are bootstrap 95\%.}
\label{fig:persistence}
\end{figure}

\subsection{Robustness}
\emph{Threshold.} The minimum of six shots per match is arbitrary; $R$ is 1.402, 1.403 and 1.397 with minima of 6, 10 and 15 shots (1{,}550, 1{,}549 and 1{,}508 matches; all $p<0.001$).

\emph{Parametric check.} The beta-binomial maximum-likelihood estimate $\hat\rho=0.0174$ (bootstrap 95\% CI $[0.0135,\,0.0210]$) excludes zero without using the permutation construction, tying the finding to a standard model class for overdispersed proportions (Cameron and Trivedi 1998).

\begin{figure}[h]
\centering
\includegraphics[width=0.55\textwidth]{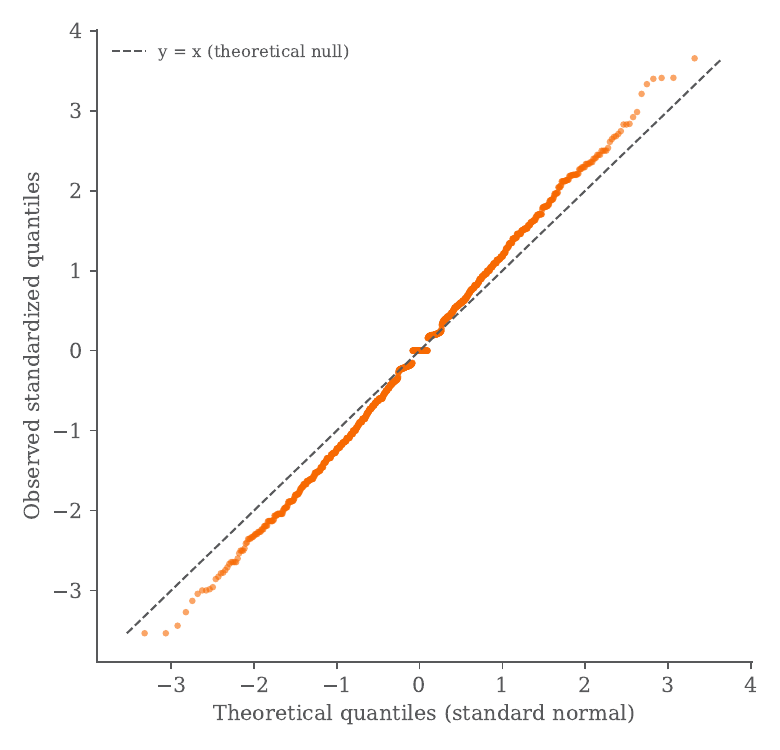}
\caption{Q--Q plot of standardized proportions against standard normal quantiles, pooled ($n=1550$). Tails bow away from the reference line in both directions, the signature of overdispersion.}
\label{fig:qq}
\end{figure}

\emph{Normal scores.} Standardized proportions $(\hat p_i-0.5)/\sqrt{0.25/n_i}$ have standard deviation 1.19 ($\approx\sqrt{1.40}$) and heavier-than-normal tails (Figure~\ref{fig:qq}).

\subsection{An independent dataset}
We applied the identical shot-based test to the FA Women's Super League 2020/21 (130 usable matches). The result is $R=1.16$ ($p=0.094$): not significant. With the sample's own shot counts and $R^\star=1.40$, Algorithm~\ref{alg:power} gives power 0.84, so the sample could have detected an effect of the men's-league size but not necessarily a smaller one. The data are compatible with a real but smaller effect and with none, and one season does not allow more.

\subsection{Two illustrative matches}
At the same ground, Celta Vigo's home matches against Real Sociedad and against Villarreal produced End~P shares of 82\% (18 shots to 4) and 18\% (3 to 14). Both matches had the same attractor share, 82\%, but opposite labels, illustrating that the attractor can be either end and that its strength can be large in a single match. These are extreme cases among 379 and are shown for illustration, not as evidence.

\subsection{Predictability within a match}
A coach might act on a busy end in the opening minutes. We test how much of the pattern in a half is predictable from an earlier part of the match. For each league, Table~\ref{tab:predictability} gives the correlation between the home-minus-away shot differential in minutes 0--23 and in the remainder of the first half, and between the two halves.

\begin{table}[h]
\centering\small
\begin{tabular}{@{}lcc@{}}
\toprule
League & Within half ($r$, 0--23 vs.\ rest of half 1) & Across halves ($D_1$ vs.\ $D_2$) \\
\midrule
La Liga        & $0.172$ ($p<0.001$) & $0.149$ ($p=0.004$) \\
Premier League & $0.105$ ($p=0.041$) & $0.303$ ($p<0.001$) \\
Serie A        & $0.165$ ($p=0.001$) & $0.271$ ($p<0.001$) \\
Ligue 1        & $-0.010$ ($p=0.85$) & $0.232$ ($p<0.001$) \\
\bottomrule
\end{tabular}
\caption{Predictability of shot-differential patterns within a half versus across halves.}
\label{tab:predictability}
\end{table}

Within-half correlations are weak (at most 0.17, and about zero in Ligue 1); in the Premier League an early differential explains about 1\% of the later one ($r^2=0.011$). The positive across-half correlations reflect persistent home-versus-away strength ($m$ in Proposition~\ref{prop:identify}), which cancels from the end split.

To turn this into a forecast, we use StatsBomb's expected-goals value to define a clear chance ($xG\ge0.1$) and predict each team's second-half clear chances in La Liga with two nested Poisson regressions compared by 5-fold cross-validation: Model~A uses the team's leave-one-out season rate and a home indicator, and Model~B adds this match's first-half clear chances and first-half $xG$.

\begin{table}[h]
\centering\small
\begin{tabular}{@{}lcc@{}}
\toprule
 & Model A (pre-match information) & Model B (+ this match's first half) \\
\midrule
Out-of-sample MAE & 1.242 & 1.249 \\
Out-of-sample log-likelihood & $-1342.2$ & $-1340.8$ \\
\bottomrule
\end{tabular}
\caption{Cross-validated accuracy for second-half clear-chance counts, La Liga 2015/16 ($n=758$ team-matches).}
\label{tab:predictive}
\end{table}

Adding the match's own first half does not improve out-of-sample accuracy. In the full-sample fit the season rate ($\hat\beta=0.301$) and home indicator ($\hat\beta=0.229$) are strongly significant ($p<0.0001$), while first-half clear chances ($\hat\beta=0.053$, $p=0.088$) and first-half $xG$ ($\hat\beta=0.031$, $p=0.727$) are not; McFadden pseudo-$R^2$ is 0.040.

\section{Discussion and Limitations}

What the evidence establishes is narrow and, we think, clear. In shot counts, the allocation of events to the two physical ends varies across matches by more than chance allows (pooled $R=1.40$; replicated in four leagues; robust to score state), so the Goal Attractor, read as a match-level asymmetry in where goalmouth action falls, exceeds what coin-flip allocation predicts. The effect is small: a typical underlying attractor share of about 55\%, and an observed excess of 1.6 percentage points over the 58\% that chance alone yields. The dispersion has similar size for shots on target and clear chances, and goals are too sparse to test.

What it does not establish is a persistent, end-linked mechanism. Proposition~\ref{prop:structure} shows the test responds to any between-halves change in relative attacking rates, and Section~3.4 shows that a persistent end effect of the size needed to explain the excess is rejected, while a modest amount of half-by-half variation in team intensity (coefficient of variation near 0.28) reproduces it. We therefore describe the finding as match-level end asymmetry beyond chance, without attributing it to the ground, the weather, or any other fixed feature of the end. Such an attribution would require data this study does not have, for example end identity linked to stadium geometry, wind, sun or pitch condition, or tracking data.

Several limitations apply. The data are one season per league from an open release, and Bundesliga is partial. The Poisson structure behind Proposition~\ref{prop:structure} is a modelling assumption, although the permutation test needs only the exchangeability it implies under $H_0$. The window contrast rests on stationary short-memory shocks, and its use as an upper bound requires $c_w\ge c_x$. Own goals are not tracked in the score-state analysis. The women's league sample is too small to settle whether the effect differs in size. Finally, the weak within-match predictability means the asymmetry is not usable as an in-game signal; a monitoring tool would need a sequential statistic with a measured false-alarm rate.

\section{Conclusion}

Nearly every football match has a busier goal, but only a small part of that is more than chance. Across 1{,}550 matches in five leagues the end split of shots is overdispersed ($R=1.40$, $\hat\rho=0.017$), replicates, survives score-state controls and appears with similar strength for shots on target and clear chances. A persistent end-linked effect does not account for it, and the asymmetry is not predictable enough within a match to act on. We provide the model, four properties of the statistic, a test and a reproducible implementation, so that richer data on stadium geometry, weather and tracking can be used to ask what makes a goal an attractor.

\section*{Data and Code Availability}
All shot-event data are from StatsBomb Open Data (StatsBomb 2024), freely available at \url{https://github.com/statsbomb/open-data} under StatsBomb's public data license. A standalone Python script implementing Algorithms~\ref{alg:endswap}--\ref{alg:power} (data retrieval, end assignment, the permutation test, bootstrap intervals, the beta-binomial fit, the outcome-specific analyses and the persistent-end contrast) is included as ancillary material and reproduces the tables in this paper from raw open data.

\section*{References}
\begingroup
\parindent0pt
\parskip8pt
\raggedright

\hangindent2em Cameron, A. Colin, and Pravin K. Trivedi. 1998. \emph{Regression Analysis of Count Data}. Cambridge: Cambridge University Press.

\hangindent2em Efron, Bradley, and Robert J. Tibshirani. 1993. \emph{An Introduction to the Bootstrap}. New York: Chapman \& Hall.

\hangindent2em Good, Phillip I. 2005. \emph{Permutation, Parametric, and Bootstrap Tests of Hypotheses}. 3rd ed. New York: Springer.

\hangindent2em Johnson, Norman L., Adrienne W. Kemp, and Samuel Kotz. 2005. \emph{Univariate Discrete Distributions}. 3rd ed. Hoboken, NJ: Wiley.

\hangindent2em Kingman, J. F. C. 1993. \emph{Poisson Processes}. Oxford: Oxford University Press.

\hangindent2em Pollard, Richard. 1986. ``Home Advantage in Soccer: A Retrospective Analysis.'' \emph{Journal of Sports Sciences} 4 (3): 237--48.

\hangindent2em Price, Kelvin, Hengrui Cai, Weining Shen, and Guanyu Hu. 2022. ``How Much Does Home Field Advantage Matter in Soccer Games? A Causal Inference Approach for English Premier League Analysis.'' arXiv preprint, arXiv:2205.07193.

\hangindent2em Qi, Minhao, Hengrui Cai, Guanyu Hu, and Weining Shen. 2025. ``Time-Varying Home Field Advantage in Football: Learning from a Non-Stationary Causal Process.'' arXiv preprint, arXiv:2506.11399.

\hangindent2em Reich, Brian J., James S. Hodges, Bradley P. Carlin, and Adam M. Reich. 2006. ``A Spatial Analysis of Basketball Shot Chart Data.'' \emph{The American Statistician} 60 (1): 3--12.

\hangindent2em StatsBomb. 2024. \emph{StatsBomb Open Data} [data set]. GitHub. \url{https://github.com/statsbomb/open-data}.

\hangindent2em Yamada, Kazuhiro, and Keisuke Fujii. 2026. ``STAMP: A Shot-Type-Aware Areal Multilevel Poisson Model for League-Wide Comparison of Basketball Shot Charts.'' arXiv preprint, arXiv:2603.24015.

\endgroup

\end{document}